\documentclass[aps,prd,twocolumn,showpacs,showkeys,nofootinbib,superscriptaddress]{revtex4-1}

\usepackage{longtable}
\usepackage{graphicx}
\usepackage{amsmath}
\usepackage{amsfonts}
\usepackage{amssymb}
\usepackage{natbib} 
\usepackage{setspace}
\usepackage{xspace}
\usepackage{cancel}
\usepackage{physics}
\usepackage{tensor,braket,color,bbm,slashed}

\usepackage{newtxtext,newtxmath}
\usepackage{mathrsfs}

\usepackage{ulem}

\begin{document}

\title{
  \begin{flushright}
    \rightline{\hfill APCTP Pre2019-012, LFTC-19-7/45} 
  \end{flushright}
  Electroweak properties of kaons in a nuclear medium}

\author{Parada~T.~P.~Hutauruk}
\email{parada.hutauruk@apctp.org}
\affiliation{Asia Pacific Center for Theoretical Physics, Pohang, Gyeongbuk 37673, Korea }

\author{K.~Tsushima}
\email{kazuo.tsushima@gmail.com}
\affiliation{Laborat\'orio de F\'isica Te\'orica e Computacional, Universidade Cruzeiro 
  do Sul / Universidade Cidade de Sao Paulo, 01506-000, S\~ao Paolo, SP, Brazil }
\affiliation{Asia Pacific Center for Theoretical Physics, Pohang, Gyeongbuk 37673, Korea }

\date{\today}

\begin{abstract}
Kaon electroweak properties in symmetric nuclear matter are
studied in the Nambu--Jona-Lasinio model using the proper-time regularization. 
The valence quark properties in symmetric nuclear matter are calculated 
in the quark-meson coupling model, and they are used as inputs for studying     
the in-medium kaon properties in the NJL model. 
We evaluate the kaon decay constant, kaon-quark coupling constant,  
and $K^+$ electromagnetic form factor by two different approaches.
Namely, by two different ways of calculating the in-medium constituent quark 
masses of the light quarks.   
We predict that, in both approaches, the kaon decay constant  
and kaon-quark coupling constant decrease as nuclear density increases, 
while the $K^+$ charge radius increases by 20-25\% at normal nuclear density.
\end{abstract}


\maketitle

\section{Introduction}    \label{intro}

Kaons play special roles in strong and electroweak interaction phenomena,
and in low-energy hadronic reactions associated with 
quantum chromodynamics (QCD) at low 
energies~\cite{Aichelin:1986ss,Tsushima:1994rj,Tsushima:1994pv,WKW96,Barth:1997mk,
Tsushima:1996xc,Sibirtsev:1997ru,Tsushima:1997df,Shyam:1999nm,Tsushima:1998jz,Laue:1999yv,
Sibirtsev:1998jf,Tsushima:2000hs,Tsushima:2000re}.
They appear as Nambu-Goldstone bosons like pions, but show the  
flavor SU(3) symmetry breaking by the heavier strange quark mass.
Thus, kaons can provide us with useful information on chiral symmetry and
its restoration, SU(3) flavor breaking effect as well as 
the nonperturbative bound state nature of QCD.
The internal structure of kaons 
can be explored by the electromagnetic form factors (EMFFs), 
which reflect the underlying quark-gluon dynamics~\cite{GCLRT17,HBCT18}. 
Many attempts have been made to understand the internal structure of kaons    
using various approaches, e.g., the Dyson-Schwinger equation (DSE), 
Nambu--Jona-Lasinio (NJL) model,
chiral perturbation theory (ChPT), and instanton 
vacuum-based models~\cite{Zovko74,BM88a,BWI95,BRT96,BT02,NK07b,RA08,DDEF12,NBC14,KTT16,HCT16,
GCLRT17,HBCT18,AH19,Hutauruk18}.
Experimentally, $K$-meson EMFFs are poorly known, except for the $K^-$ 
in the low $Q^2$ region ($Q^2 < 0.2$ GeV$^2$)~\cite{ABBB86b}, 
where $Q^2 = -q^2$ and $q$ is the four-momentum transfer.
Recently, $K^+$ EMFF was extracted from the kaon electroproduction data 
at JLab up to $Q^2 = 2.07$ GeV$^2$~\cite{FPI-2-18}, but the data uncertainties
are still large. We await the new coming data with better precision 
up to $Q^2 \simeq 5.5$ GeV$^2$.

On the other hand, theoretical studies for $K$-meson properties  
have been made mostly in vacuum, but not in medium.
For example, kaon properties and $K^+$ EMFF in symmetric nuclear matter 
were studied only recently in Ref.~\cite{YDDTF17} based on 
a light-front constituent quark model.
There, the in-medium valence light-quark properties were calculated in 
the quark-meson coupling (QMC) model, and they were used as inputs to study 
the in-medium kaon properties.
Although this study provides us with some insights on the kaon properties  
in a nuclear medium, more complete studies are necessary for the following reasons;  
(i) the constituent quark masses in vacuum are the input parameters
in Ref.~\cite{YDDTF17}, but it is preferable to calculate them dynamically, 
(ii) the vacuum in the light-front approach is generally believed to be "trivial",
and there is no clearly defined quark chiral condensates in the 
light-front approach. 
Thus, the model does not have the dynamical chiral symmetry breaking mechanism 
nor does the model have direct connection with the emergence of
(pseudo-)Goldstone bosons such as kaons, 
and (iii) the in-medium kaon decay constant as well as  
the kaon-quark coupling constant are assumed to be density-independent. 
The purpose of the present work is to improve further the work of Ref.~\cite{YDDTF17}.
We extend the approach used for studying the pion properties 
in symmetric nuclear matter in Ref.~\cite{HOT18b}, and study the in-medium kaon 
properties using the Nambu--Jona-Lasinio (NJL) model~\cite{NJL61a,NJL61b}, 
supplemented by the QMC model inputs. 
The NJL model is a powerful chiral effective quark theory of 
low-energy QCD. Importantly, the model describes the dynamical chiral symmetry  
breaking, the origin of the pseudoscalar Goldstone bosons such as pions and kaons.  
Furthermore, the model satisfies the chiral limit as QCD dictates.
Thus, the NJL model, which has several improved aspects as 
addressed above, is suitable to study the in-medium kaon properties.

The experimental evidences, such as the EMC effect~\cite{EMC83b,Geesaman:1995yd} 
and the observed modifications of bound proton EMFFs 
at JLab~\cite{SDAA02}, suggest 
that the internal structure of hadrons would be modified in a nuclear medium. 
The phenomena of in-medium modifications of hadron 
properties~\cite{BR91,HH08,LMM09,TSTW97,STT05,KTT17,HHSYY16,MNB17,HB05,LTTWS98,RCG89,VW91} 
are tightly connected with partial restoration of chiral 
symmetry~\cite{BR96,HR16,JHK00b,JHK07}. 
The order parameters of chiral symmetry in QCD are the light-quark chiral condensates, 
and their changes are expected to be one of the most important driving forces  
for the change of hadron properties in a nuclear medium.

Spontaneous breaking of chiral symmetry generates the nonet of massless 
pseudoscalar Goldstone bosons. But the explicit breaking of U(1) axial symmetry 
selectively shifts up the $\eta^0$-singlet mass, leaving the SU(3)
flavor octet of pions, $K$-mesons, and $\eta^8$ to be massless.
Then the explicit chiral symmetry breaking by non-vanishing current quark 
masses leads to the experimentally observed 
low-lying pseudoscalar meson mass spectrum~\cite{BM88,KLVW90}.
Since the chiral symmetry has such a big impact on 
the low-lying pseudoscalar meson mass spectrum with its explicit breaking,
partial restoration of chiral symmetry 
in a strongly interacting medium is inevitable to study   
the change of hadron properties in a nuclear medium.
Studying of kaon property changes in medium
allows us to explore the nature of the bound state with 
the light and strange quarks in different environment from vacuum.

In the present work, we study the spacelike $K^+$-meson EMFF,
kaon leptonic (weak) decay constant, 
and kaon-quark coupling constant in symmetric nuclear matter using the NJL model.
The NJL model has been very successful in studying
pseudoscalar meson properties~\cite{BM88,KLVW90,BJM88}.
Recently, the model was applied to study the $K^+$ EMFF in 
vacuum~\cite{HCT16,NBC14,CBT14},
kaon valence quark distributions in medium~\cite{HMOT19},
and in-medium $\pi^+$ EMFF~\cite{HOT18b}.
In these works the in-medium valence light-quark properties    
were calculated in the QMC model, and they were used as inputs 
for studying the in-medium pion and kaon properties in the NJL model.
Extending the works of Refs.~\cite{HCT16,HOT18b}, 
we study here the electroweak properties of kaons in symmetric nuclear matter 
using the NJL model.

For studying the in-medium kaon properties microscopically,  
we need the valence quark properties in a nuclear medium. 
These are calculated by the QMC model, and used as inputs for studying
the in-medium kaons properties.
For this purpose, we adopt two approaches in this work.  
The first approach is that, the in-medium constituent quark masses
are calculated by the QMC model~\cite{STT05},
which will be denoted by ``QMC-based approach''.
The second approach is that, the in-medium constituent quark masses 
are calculated by solving the NJL model gap equations~\cite{Klevansky92} 
using the in-medium ``current quark properties'' calculated by the QMC model
as inputs.
This will be denoted by ``NJL-based approach''.
Although the two approaches yield different
in-medium constituent quark masses of the light quarks,
they lead to very similar predictions for the $K^+$ EMFF and its charge radius 
in symmetric nuclear matter.

This paper is organized as follows. 
In Sec.~\ref{sec:kaon}, we present the in-medium constituent 
valence-quark and kaon properties based on the two approaches, 
the QMC-based and NJL-based approaches.
In section~\ref{mediumFormFactors} we present how  
the in-medium $K^+$ EMFF in the NJL model is calculated, while the  
numerical results are given in Sec.~\ref{results}.
We summarize and conclude in section~\ref{summary}.

\section{In-medium kaon properties} \label{sec:kaon}

In this section we discuss the details of the quark and kaon properties in
symmetric nuclear matter 
with both the QMC-based and NJL-based approaches.
To study the $K^+$ EMFF in the NJL model, we need the in-medium valence quark 
properties as inputs calculated in the QMC model. 
There are two ways of adapting the inputs.
It is concerned about how the in-medium constituent quark 
masses are calculated and used. 
Namely, we first evaluate the in-medium ``current quark'' properties in the QMC model 
and then calculate the in-medium dynamical (constituent) quark masses 
in the NJL model using the QMC model inputs. 
This approach is called the NJL-based approach.
A similar approach was adopted in the study of the in-medium 
pion properties in Ref.~\cite{HOT18b}.
The second approach is called the QMC-based approach, uses the constituent 
quark mass values in vacuum determined in the NJL model, 
and the values are plugged in the QMC 
model to calculate the density dependence of the constituent quark 
masses.

\subsection{In-medium quark properties in the QMC model}
\label{mediumQMCmodel}

We briefly review the in-medium properties of the valence  
light and strange quarks in the QMC model~\cite{Guichon88}.
The QMC model has been successfully applied for many topics of nuclear and hadronic physics.
(See, for example, Refs.~\cite{STT05,GST18}, and references therein.) 
It has also been used to study the medium modifications of the 
nucleon weak and electromagnetic form factors  
for the neutrino scattering in dense matter~\cite{HOT18,Hutauruk:2019ptu}.
Recently, the model has been further applied for studying the 
in-medium properties of low-lying strange, charm, and bottom 
baryons~\cite{Tsushima18}.

In the QMC model medium effects arise from self-consistent exchange 
of the Lorentz-scalar ($\sigma$) and Lorentz-vector ($\omega, \rho$) meson fields 
that couple directly only to the light quarks confined in hadrons. 
The physics behind this is that the light-quark chiral condensates in medium  
are more sensitively change than those of the strange and heavier quarks.
Here, we work with symmetric nuclear matter in its rest frame  
in the Hartree mean field approximation.
(For the Fock term effects in the QMC model, see Ref.~\cite{KTT98}.)

Effective Lagrangian for symmetric nuclear matter in the QMC model 
is given by~\cite{GSRT95,GTT07}
\begin{align}
  \label{eqintro1}
  \mathscr{L}_{\textrm{QMC}} & =  \bar{\psi}_N \left[ i\gamma \cdot \partial - M_{N}^{*}(\sigma)
    - g_\omega^{} \omega^{\mu} \gamma_\mu \right] \psi_N + \mathscr{L}_\textrm{meson}\, ,
\end{align}
where $\psi_N$, $\sigma$, and $\omega$ are respectively  
the nucleon, scalar meson $\sigma$, and vector meson $\omega$ fields.
Note that the isospin-dependent $\rho$-meson field is absent, 
since we use the Hartree approximation for symmetric nuclear matter (isospin saturated), 
and the $\rho$ mean field vanishes.
The effective nucleon mass $M_N^{*} (\sigma)$ is defined by
\begin{align}
  \label{eqintro2}
  M_{N}^{*} \left( \sigma \right) &= M_{N}   - g_\sigma^{} ( \sigma ) \sigma.
\end{align}
Here, $g_{\sigma}^{} ( \sigma )$ and $g_{\omega}^{}$ 
are the $\sigma$-dependent nucleon-$\sigma$ coupling strength and
the nucleon$-\omega$ coupling constant, respectively. 
We define $g^N_\sigma \equiv g_\sigma (\sigma = 0)$ for later convenience.
The mesonic Lagrangian density $\mathscr{L}_\textrm{meson}$ in Eq.~(\ref{eqintro1})  
is given by 
\begin{eqnarray}
  \label{eqintro3}
  \mathscr{L}_\textrm{meson} &= & \frac{1}{2} (\partial_\mu \sigma \partial^\mu 
\sigma - m_\sigma^2 \sigma^2)
  - \frac{1}{2} \partial_\mu \omega_\nu (\partial^\mu \omega^\nu - \partial^\nu \omega^\mu)
  \nonumber \\ &&\mbox{} 
  + \frac{1}{2} m_\omega^2 \omega^\mu \omega_\mu.  
\end{eqnarray}

In the Hartree approximation for symmetric nuclear matter, the nucleon Fermi momentum 
$k_F$ is related with the baryon ($\rho_B$) and scalar ($\rho_s$) densities
respectively, 
\begin{align}
  \label{eqintro4}
  \rho_{B}^{} &= \frac{4}{(2\pi)^3} \int d{\bf k}\, 
\theta \left( k_F - | {\bf k} | \right) =
  \frac{4 k_F^3}{3 \pi^2}, \nonumber \\
  \rho_{s}^{} &= \frac{4}{(2\pi)^3} \int d {\bf k}\,  
\theta \left( k_F - | {\bf k} | \right)
  \frac{M_N^{*} (\sigma)}{\sqrt{M_N^{*2} (\sigma ) + {\bf k}^2}}.
\end{align}

In the QMC model~\cite{STT05,GSRT95,GTT07}, nuclear matter is treated 
as the collection of nonoverlapping nucleon-MIT bags~\cite{CJJTW74}.
The Dirac equations for the quarks and antiquarks in a hadron bag are given by
\begin{align}
  \label{eqintro5}
  \left[ i \gamma \cdot \partial_{x} - \left( m_l - V_{\sigma}^{l} \right) \mp \gamma^{0} 
  \left( V_{\omega}^{l} + \frac{1}{2} V_{\rho}^{l} \right) \right] \left( \begin{array}{c}
    \psi_u(x)  \\ \psi_{\bar{u}}(x) \\ \end{array}
  \right) &= 0, \nonumber \\
  \left[ i \gamma \cdot \partial_{x} - \left( m_l - V_{\sigma}^{l} \right) \mp \gamma^{0} 
  \left( V_{\omega}^{l} - \frac{1}{2} V_{\rho}^{l} \right) \right] \left( \begin{array}{c}
    \psi_d(x)  \\ \psi_{\bar{d}}(x) \\ \end{array} \right) &= 0,  \nonumber \\
  \left[ i \gamma \cdot \partial_{x} - m_{s} \right] \left( \begin{array}{c}
    \psi_s(x)  \\ \psi_{\bar{s}}(x) \\ \end{array} \right) &= 0,
\end{align}
where $l$ stands for ``light'', i.e., $l = u$ or $d$, 
and the effective light-quark mass $m_l^{*}$ is defined as
\begin{align}
  \label{eqintro5a}
  m_l^{*} & \equiv m_l - V_{\sigma}^{l}.
\end{align}
Here we assume SU(2) symmetry, $m_l = m_u = m_d$ for the valence quark masses 
throughout this study 
(thus $m^*_l = m^*_u = m^*_d$), 
and $m_s^{}$ is the strange valence quark mass in vacuum, 
and $-V_\sigma^{l}$ is the Lorentz scalar potential, which couples only to 
the light quarks. The strange quark is decoupled from the scalar and vector potentials  
so that its effective mass does not change in a nuclear medium, $m_s^* = m_s$.

The scalar and vector mean fields in symmetric nuclear matter are defined, respectively 
by the mean expectation values,  
\begin{eqnarray}
V_{\sigma}^{l} \equiv g_{\sigma}^{l} \sigma = g_{\sigma}^{l} \braket{\sigma},
\qquad
V_{\omega}^{l} \equiv g_{\omega}^{l} \omega = g_{\omega}^{l} \delta^{\mu,0} 
\braket{ \omega^{\mu} },
\end{eqnarray}
where the light-quark-meson coupling constants, $g_{\sigma}^{l}$ and 
$g_{\omega}^{l}$, are defined later through Eq.~(\ref{Ssigma}).
The eigenenergies in units of $1/ R_h^{*}$
with the bag radius of the hadron $h$, are given by  
\begin{align}
  \label{eq:kaonmed9}
  \left( \begin{array}{c}
    \epsilon_u \\
    \epsilon_{\overline{u}}
  \end{array} \right)
  & = \Omega_l^* \pm R_h^* \left( V^l_\omega + \frac{1}{2} V^l_\rho \right), \nonumber \\
  \left( \begin{array}{c} \epsilon_d \\
    \epsilon_{\overline{d}}
  \end{array} \right)
  &= \Omega_l^* \pm R_h^* \left( V^l_\omega - \frac{1}{2} V^l_\rho \right), \nonumber \\
  \epsilon_{s}^{} &= \epsilon_{\overline{s}}^{} = \Omega_{s}^*.
\end{align}
where 
\begin{eqnarray}
\Omega^{*}_l &=& \Omega^{*}_{\overline{l}} = \left[ x_l^2 + \left(R_h^{*}  m_l^{*} 
\right)^2 \right]^{1/2},
\nonumber \\
m_l^{*} &=& m_l^{} - V^l_\sigma\, = m_l - g_\sigma^{l} \sigma , 
\nonumber \\
\Omega_{\overline{s}}^{*} &=& \Omega_{s}^{*} = [x_{s}^2 + (R_h^{*} m_s )^2 ]^{1/2}.
\end{eqnarray}
The effective mass of hadron $h$ in nuclear medium $m_h^{*}$, which will be shown
to be Lorentz scalar, is calculated as
\begin{align}
  \label{eq:kaonmed10}
  m_h^{*} &= \sum_{j = l, \bar{l}, s, \bar{s}} \frac{n_j \Omega_j^{*} -z_h}{R^{*}_h} 
+ \frac{4}{3} \pi R_h^{* 3} B,
\end{align}
and the in-medium bag radius $R_h^{*}$ is determined by the stability condition 
for a given baryon density self-consistently, 
\begin{align}
  \frac{d m_h^{*}}{d R^*_h}  = 0. 
\end{align}
In Eq.~(\ref{eq:kaonmed10}) $z_h$ is assumed to be density independent.
It is related with the sum of the center-of-mass and 
gluon fluctuation corrections~\cite{GSRT95}, which is determined by 
the hadron mass in vacuum.
The bag constant $B$ is fixed by the inputs 
for the nucleon in vacuum, namely, $R_N = 0.8$~fm with $M_N^{} = 939$~MeV, 
for a chosen value of $m_l$.

The ground state wave function of a quark in hadron $h$ satisfies 
the boundary condition at the bag surface,
\begin{equation}
j_0^{} (x) =  \beta_{q} j_1^{} (x)
\end{equation} 
with $q = l$ ($=u,d$) or $s$, and $j_0^{}$ and $j_1^{}$ are the 
spherical Bessel functions, and 
\begin{align}
  \label{eq:beta}
\beta_{q} &= \sqrt{\frac{\Omega_{q}^{*} - m_{q}^{*} R_h^{*}}{\Omega_{q}^{*} 
+ m_{q}^{*} R_h^{*}}}.
\end{align}
The scalar $\sigma$ and vector $\omega$ meson fields at the nucleon 
level are related as
\begin{align}
  \label{eq:kaonmed11}
  \omega &= \frac{g_\omega^{} \rho_B^{} }{m_\omega^2}, \nonumber \\
  \sigma &= \frac{4 g_\sigma^{N} C_N (\sigma)}{(2\pi)^3m_{\sigma}^2}
\int d {\bf k}\, \theta (k_F - | {\bf{k}} | ) \frac{M_N^{*} (\sigma)}{\sqrt{M_N^{*2} 
(\sigma) + {\bf{k}}^2}},
\end{align}
where $C_N (\sigma)$ is defined by
\begin{align}
  C_N (\sigma) &= \frac{-1}{g_\sigma^{N}} 
\left( \frac{\partial M_N^{*} (\sigma )}{\partial \sigma } \right), 
\end{align}
with $C_N (\sigma) = 1$ for the pointlike nucleon case such as 
in Quantum Hadrodynamics (QHD)~\cite{SW86,SW97}.
The $\sigma$-dependent coupling $g_\sigma (\sigma)$ or $C_N (\sigma)$ 
is the origin of the novel saturation properties in the QMC model, and
the quark dynamics is included in the effective nucleon 
mass $M_N^{*} (\sigma$) via a self-consistent manner as
in Eqs.~(\ref{eq:kaonmed10}) [with $m^*_h \to M^*_N$] and~(\ref{eq:kaonmed11}).
By solving the self-consistent equation for the $\sigma$ mean field    
Eq.~(\ref{eq:kaonmed11}), the total energy per nucleon is calculated by
\begin{eqnarray}
  \label{eq:kaonmed12}
  E^{\rm tot}/A &=& \frac{4}{(2\pi)^3 \rho_B}
  \int d {\bf k}\, \theta (k_F - | {\bf{k}} |) \sqrt{M_N^{*2} (\sigma)  + {\bf{k}}^2} 
    \nonumber \\ &&  
  + \frac{m_\sigma^2 \sigma^2}{2\rho_B^{}} + \frac{g_\omega^2 \rho_B^{}}{2 m_\omega^2}.
\end{eqnarray}

The coupling constants $g^N_\sigma ( = g_\sigma(\sigma=0))$ 
and $g_\omega = g^N_\omega$  
in Eq.~(\ref{eq:kaonmed12}) ($g^N_\sigma$ is implicit) are determined to reproduce 
the binding energy of symmetric nuclear matter 15.7 MeV 
at the saturation density $\rho_0^{} =  0.15$ fm$^{-3}$, and they are, respectively, 
$g_\sigma^{N} = 3 g^l_{\sigma} S_N (\sigma = 0)$
and $g_\omega^{} = 3 g_\omega^l$. In this study, we will use $m_l = 16.4$ and 
$400$ MeV, and thus we give the explicit values of the coupling constants for 
the both quark mass values. Namely, $ g_\sigma^l\simeq 5.6251~[4.7009]$
for $m_l = 16.4~[400]$~MeV, where $S_N(\sigma)$ is defined 
through~\cite{Guichon88,KTT17}
\begin{eqnarray}
\dfrac{d M_{N}^*(\sigma)}{d \sigma}
&=& - 3 g_{\sigma}^l \int_{\rm bag} d^3 y \, \bar{\psi}_l({\bf y})~\psi_l({\bf y})
\nonumber \\
&\equiv& - 3 g_{\sigma}^l S_{N}(\sigma) = - \dfrac{d}{d \sigma} 
\left[ g^{}_\sigma(\sigma) \sigma \right], 
\label{Ssigma}
\end{eqnarray}
which gives $S_N(\sigma=0) = 0.4899~[0.6950]$ for $m_l = 16.4~[400]$~MeV
and $R_N = 0.8$ fm with $\psi_l$ being the lowest mode bag wave function in medium.
Note that, the right hand side of Eq.~(\ref{Ssigma}) is the quark scalar charge, 
which is Lorentz scalar, and thus the left-hand-side of Eq.~(\ref{Ssigma}) 
is Lorentz scalar, and thus $M^*_N(\sigma)$ as well.
These relations show that the in-medium quark dynamics is  
explicitly included in the QMC model.
Since the light quarks in any hadrons should generally feel the same scalar 
and vector mean fields as those in the nucleon
(after one chooses a fixed light-quark mass value in vacuum), 
we can systematically study the hadron properties in medium 
without introducing any new coupling constants for   
the $\sigma$ and $\omega$ mean fields for different hadrons.

\subsection{Kaon properties in the NJL model} 
\label{mediumNJL}

Based on the SU(3) NJL model in vacuum described 
in Ref.~\cite{HCT16}, we explore the in-medium properties of dynamical 
quark and kaon properties in this section.
The medium effect is implemented through the in-medium  
potentials for the valence quarks generated by the QMC model.
The dynamical quark mass in medium, $M^*_q$ ($q=u,d,s$),
in the NJL model with the proper-time regularization 
scheme is given by~\cite{HOT18b}
\begin{align}
  \label{eq:kaonmed13}
  M_q^{*} & = m_q^{*} + \frac{3G_\pi M_q^{*}}{\pi^2} \int_{1/\Lambda_{\rm UV}^2}^{\infty} 
  \frac{d\tau}{\tau^2} e^{-\tau M_q^{*2}},
\end{align}
where $M_q^{*}$ and $m_q^{*}$ are the in-medium dynamical 
and current quark masses, respectively. 
$G_\pi$ is the four-fermion coupling constant, which is taken the same as in vacuum, 
and $\Lambda_{\rm UV}$ is the ultraviolet cutoff. For the infrared cutoff 
$\Lambda_{\rm IR}$, we take $\tau_{\rm IR} = 1/\Lambda_{\rm IR}^2 = + \infty$ in nuclear medium. 
(This change in $\Lambda_{\rm IR}$ from that in Ref.~~\cite{HCT16} 
gives negligible difference in results.) 
Note that the expression of $M_q^{*}$ is different from that in 
Refs.~\cite{Maedan09,WCT16}, since the information of ``nucleon'' Fermi momentum 
and nuclear matter saturation properties are included in $m^*_q$ (and $V^l_\omega$) 
calculated in the QMC model, and we should drop the density dependent term.

The in-medium dressed quark propagators in Eq.~(\ref{eq:kaonmed13}) are given by  
\begin{align}
  \label{eq:kaonmed14}
  S_l^{*}(k^*) &= \frac{\slashed{k}^* + M_l^{*}}{(k^*)^2 
- (M_l^{*})^2 + i \epsilon}, \nonumber \\
  S_s^{*}(k^*) &= S_s (k) = \frac{\slashed{k} + M_s}{k^{2} - M_s^2 + i \epsilon},
\end{align}
where the quantities with the asterisk above denote those in medium,  
and medium effects enter in the light-quark mass $M^*_l$ and  
momentum $k^\mu$ by $k^{* \mu} = k^\mu + V^\mu$ 
due to the vector mean field $V^\mu = (V^0, \textbf{0})$. 
The modification of the space component of the light-quark momentum
$k^{* \mu}$ is neglected, since it is known very small~\cite{KTT98}.

In Eq.~(\ref{eq:kaonmed13}) the vector potential entering in the 
light-quark propagator was eliminated by the shift of the integral variable~\cite{CMPR09}.
Note that, the in-medium strange quark propagator is the same as that in 
vacuum, since the strange quark is decoupled from the scalar 
and vector mean fields in the QMC model, and the same assumption is adopted. 
Namely, $M_s^* = M_s$ is assumed also in the present NJL model calculation.

The description of kaon as the dressed quark-antiquark bound state 
is obtained by solving the Bethe-Salpeter equation (BSE)
in the random phase approximation (RPA).
The solution to the BSE in each meson channel is given by the two-body 
scattering $t$-matrix, that depends on the interaction channel. 
We also introduce below the reduced $t$-matrix 
for vector mesons, since it will be used later in Eqs.~(\ref{eq:f1U}) and (\ref{eq:f1S}). 
The reduced $t$-matrices in the $K$-meson and vector meson $V$ take the forms~\cite{HCT16},

\begin{align}
  \label{eq:tmatrix}
  \tau_K^{*} (p^*) &= \frac{-2i\,G_\pi}{1 + 2\,G_\pi\,\Pi^{*}_K (p^{*2})}, \nonumber \\
  \tau_V^{* \mu \nu} (p^*) &= \frac{-2i\,G_\rho}{1 + 2\,G_\rho\,\Pi^{*}_V (p^{*2})}
  \left(g^{\mu \nu} + 2\,G_{\rho}\,\Pi^{*}_V (p^{*2})\, \frac{p^{*\mu} 
    p^{*\nu}}{p^{*2}} \right),
\end{align}
where $G_\rho$ is the four-fermion coupling constant for 
$\rho$ channel, and the bubble diagrams in nuclear medium give
\begin{align}
  \label{eq:bubblegraphtot}
  \Pi^{*}_{K} (p^{*2}) &= 6i \int \frac{d^4k}{(2\pi)^4}\,
  \mathrm{Tr}_D \left[ \gamma_5 S^{*}_{\,l}(k^*) \gamma_5 S^{*}_{\,s}(k^* + p^*) \right], 
\nonumber \\
  \Pi_{V}^{* qq} (p^{*2})\, P_T^{* \mu\nu} &= 6i\int \frac{d^4 k}{(2\pi)^4} \,
  \mathrm{Tr}_D \left[ \gamma^\mu S^{*}_{\,q} (k^*) \gamma^\nu S^{*}_{\,q}(k^* + p^*) \right],
\end{align}
with $\Pi^{*}_\rho = \Pi^{*}_\omega = \Pi_{v}^{* ll}$, and $\Pi^{*}_\phi = \Pi_{v}^{* ss}$.
The trace is only for the Dirac indices, and $P_T^{* \mu\nu} = g^{\mu\nu} 
- p^{*\mu} p^{*\nu}/p^{*2}$.

The in-medium meson mass is defined by the pole in the corresponding 
$t$-matrix as in the vacuum case, 
\begin{align}
  1 + 2\, G_\pi\, \Pi^{*}_{K} (p^{* 2} = m_{K}^{* 2}) &= 0,
\end{align}  
where the similar conditions determine other meson masses in medium. 
This leads to the in-medium kaon mass, 
\begin{align}
  m_{K}^{*2} &= \left( \frac{m^{*}_s}{M^{*}_s} + \frac{m_l^{*}}{M^{*}_l} \right)
  \frac{1}{G_\pi\, \mathcal{I}_{l s}(m_K^{* 2})} + (M^{*}_s-M^{*}_l)^2,
\end{align} 
where $m_l^* = m_u^* = m_d^*$ and $\mathcal{I}_{ls}(p^{*2})$ is defined by, 
\begin{align}
  \mathcal{I}_{ab}(p^{*2}) &= \frac{3}{\pi^2} \int_0^1 dz \int \frac{d\tau}{\tau} \nonumber \\
  & \mbox{} \times e^{-\tau\left [-z(1-z)\,p^{*2} + 2pV_0z(1-z)-z(1-z)V_0^2 
+ x\,M_b^{* 2} + (1-z)\,M_a^{* 2}\right] }.
\end{align} 
This demonstrates the Goldstone boson nature of the kaon in the chiral limit.
The residue at the pole in the $\bar{q}q$ $t$-matrices defines 
the kaon-quark coupling constant ($g_{K q q}^{*}$) in medium:
\begin{align}
  \label{eq:couplinconstant}
   (g_{K q q}^*)^{-2} &= -\left. \frac{\partial\, 
\Pi_K^* (p^{*2})}{\partial p^{*2}} \right|_{p^{*2} = m_K^{*2}}. 
\end{align}

Following Refs.~\cite{HCT16,HOT18,CBT14}, we have chosen $\Lambda_{\rm IR} =$ 240 MeV, 
which is of the order of $\Lambda_{\rm QCD}$, and $M_l =$ 400 MeV in vacuum. 
These values were fixed to give $m_\pi =$ 140 MeV and $m_K =$ 495 MeV, 
together with the pion decay constant $f_\pi =$ 93 MeV. 
Furthermore, this gives $\Lambda_{\rm UV} =$ 645 MeV, $G_\pi =$ 19.0 GeV$^{-2}$. 
Fit to the physical masses of the vector mesons, $m_\rho =$ 770 MeV 
and $m_\omega =$ 782 MeV, gives $G_\rho =$ 11.0 GeV$^{-2}$, and  
$G_\omega =$ 10.4 GeV$^{-2}$, respectively. 
These parameters are used in the present study,
except for $(1/\Lambda^2_{\rm IR}) \to \infty$ in medium.

\subsection{QMC-based and NJL-based approaches}

As mentioned already, we adopt two approaches to estimate the in-medium 
constituent quark masses.
In the NJL-based approach we use the in-medium ``current quark'' properties 
evaluated by the QMC model, 
and calculate the in-medium dynamical (constituent) 
quark masses in the NJL model.
While in the QMC-based approach, we use the vacuum constituent quark mass values
fixed by the fit in the NJL model, and the QMC model calculates 
the density dependence of the constituent quark masses.
Note that, the in-medium kaon properties including effective kaon mass $m^*_K$ 
and quark EMFFs, are all calculated in the NJL model for both approaches.

We first discuss the parameters of the NJL-based approach. 
Following Ref.~\cite{HOT18b}, we adopt $m_l^{} = m_u^{} = m_d^{} = 16.4$~MeV 
and $m_s^{} = 356$~MeV, for the vacuum current quark mass values.
These values are optimally fixed by the NJL model in Refs.~\cite{HCT16,HBCT18}. 
The other parameters in the QMC model are fixed by the nucleon mass $M_N = 939$ MeV, 
nucleon bag radius $R_N = 0.8$~fm, 
and the nuclear matter binding energy per nucleon of 15.7~MeV   
at the saturation density $\rho_0^{} = 0.15$ fm$^{-3}$.
For the QMC-based approach, the NJL-model vacuum constituent quark mass values are used 
as inputs to calculate the density dependence of the constituent quark masses 
in the QMC model. 
In this case we use $M_l = M_u = M_d = 400$~MeV and $M_s = 611$~MeV. 
They are summarized in Table~\ref{tab:model1}.

\begin{table}[htb]
\caption{
  Parameters in the QMC model and some quantities obtained at saturation density $\rho_0=$ 0.15 $\rm{fm}^{-3}$ for the 
two approaches, the NJL-based and QMC-based. 
They respectively correspond to the quark masses in vacuum, 
$m_l = $ 16.4 (NJL-based), and 400~MeV (QMC-based). 
The light-quark mass $m_l$, effective nucleon mass $M_N^{*}$, nuclear 
incompressibility $K$, and $B^{1/4}$ are given in units of MeV. 
The other parameters are obtained by using the inputs for the nucleon bag radius 
$R_N = 0.8$~fm in vacuum, nucleon mass in vacuum $M_N = 939$ MeV, 
and the nuclear matter binding energy of 15.7~MeV at the
saturation density $\rho_0$.
}
\label{tab:model1}
\addtolength{\tabcolsep}{3.4pt}
\begin{tabular}{ccccccc} 
\hline \hline
$m_{l}$ & $g_{\sigma}^2 / 4 \pi$ & $g_{\omega}^2 / 4 \pi$ & $B^{1/4}$ & $z_N^{}$ & $M_N^{*}$ & $K$ 
\\[0.2em] 
\hline
$16.4$   & $5.438$ & $5.412$ & $169.2$ & $3.334$ & $751.95$ & $281.50$ \\
$400$    & $7.645$ & $10.136$& $98.5$  & $5.321$ & $625.98$ & $339.10$  
\\ \hline \hline
\end{tabular}
\end{table}

\begin{figure}[t]
  \centering\includegraphics[width=\columnwidth]{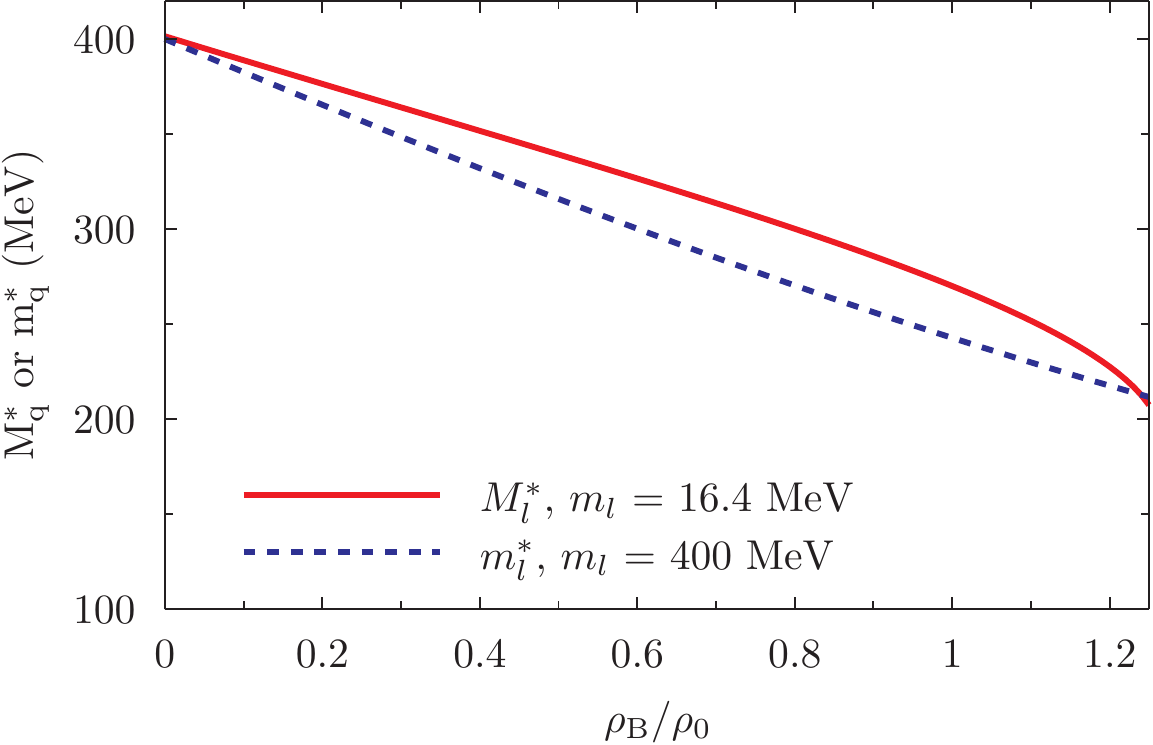}
  \caption{\label{fig:mq*} 
    In-medium constituent quark mass $M^*_l$ in the NJL-based approach
    (solid line) with $m_l =$ 16.4 MeV, and $m_l^*$ 
    in the QMC-based approach (dashed line) with $m_l =$ 400 MeV.}
\end{figure}

\begin{figure}[t]
  \centering\includegraphics[width=\columnwidth]{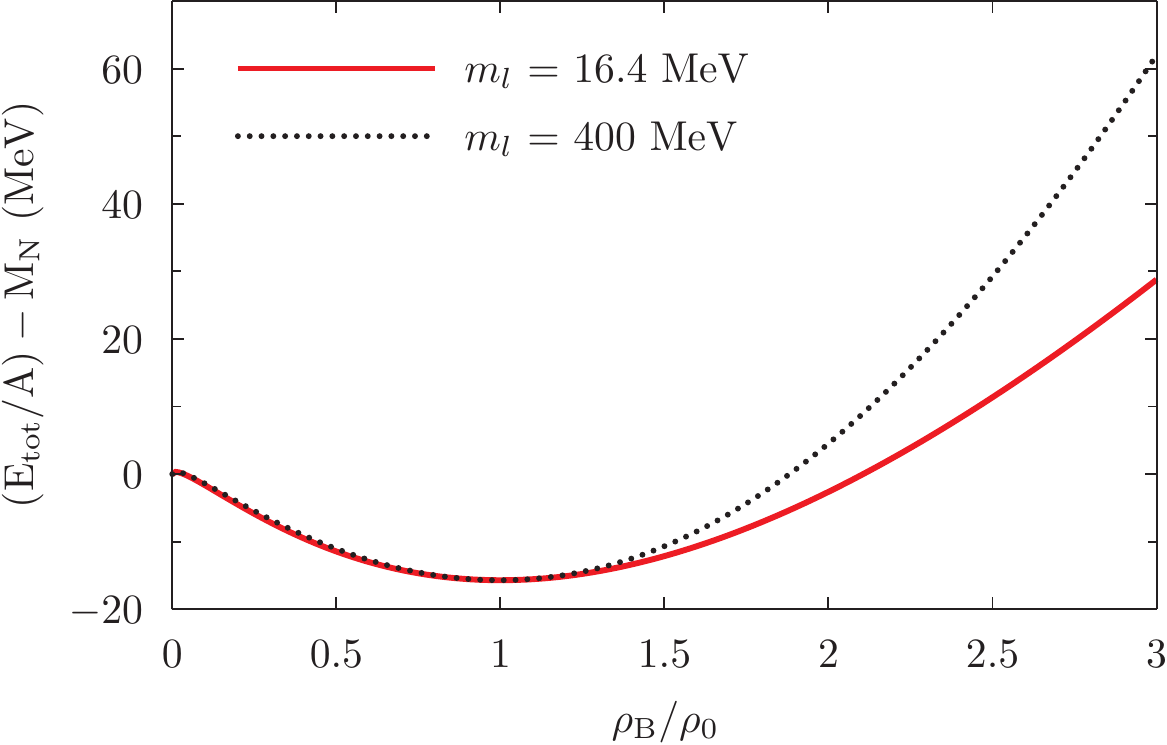}
  \caption{\label{fig1} 
    Negative of binding energy per nucleon ($E_{\rm tot} /A - M_N $) 
    for symmetric nuclear matter calculated in the QMC model for the light-quark mass 
    $m_l =$ 16.4 MeV (solid line) and that for $m_l =$ 400 MeV (dotted line).}
\end{figure}

In Fig.~\ref{fig:mq*} we compare the in-medium constituent quark masses calculated 
in the two approaches. The solid line is the $M_l^*$ ($m_l = 16.4$ MeV) 
in the NJL-based approach, 
while the dashed line is the $m_l^*$ ($m_l = 400$ MeV) in the QMC-based approach.
This shows the two approaches give similar density dependence for  
the constituent quark mass of the light-quark. The largest
difference between the two approaches is about 10\%.
Relative to the vacuum value, the light-quark constituent mass is  
reduced by 30--35\% at nuclear matter saturation density $\rho_0$.
The negative of binding energy per nucleon ($E_{\rm tot} /A - M_N $) 
for symmetric nuclear matter is also shown in Fig.~\ref{fig1} for the two approaches.

Results for the density dependence of the light-quark chiral condensate are given in 
Table~\ref{tab:model4} in the two approaches. 
We find that the in-medium light-quark chiral condensate decreases about 13--16\% 
relative to that in vacuum at normal nuclear density $\rho_0^{}$,  
and the difference in the two approaches is small.

\begin{table}[htb]
  \caption{In-medium light-quark chiral condensates, 
    $-\braket{\overline{l}\, l}^{1/3} = -\braket{\overline{u} u}^{1/3}$ (GeV),
    calculated in the two approaches.}
  \label{tab:model4}
  \addtolength{\tabcolsep}{6pt}
  \begin{tabular}{ccc} 
    \hline \hline
    $\rho_B^{} / \rho_0^{}$ & NJL-based approach & QMC-based approach  \\[0.2em]  \hline
    $0.00$  & $0.171$ & $0.171$ \\ 
    $0.25$ & $0.167$ & $0.165$ \\
    $0.50$ & $0.162$ & $0.158$ \\
    $0.75$ & $0.156$ & $0.151$ \\
    $1.00$ & $0.149$ & $0.144$ \\
    $1.25$ & $0.136$ & $0.137$
    \\ \hline \hline
  \end{tabular}
\end{table} 

In Tables~\ref{tab:model5} and \ref{tab:model2} 
we list the density dependence
of the kaon properties including the kaon-quark coupling constant
calculated in the two approaches.
We find that the in-medium light-quark dynamical quark mass, effective kaon mass, 
and kaon-quark coupling constant $g_{Kqq}^{*}$ decrease in medium 
as nuclear density increases.
The in-medium kaon leptonic decay constant $f^*_K$ decreases as nuclear density increases,  
but the reduction rate is smaller than that of pion~\cite{HOT18b}, 
which is consistent with the result in Ref.~\cite{BM88d}.

\begin{table}[htb]
\caption{Constituent quark mass of the light quark and kaon properties in symmetric nuclear matter calculated in the NJL-based approach ($m_l = 16.4$ MeV). 
The masses and decay constant are given in units of GeV while  
the coupling constant $g_{Kqq}^*$ is dimensionless.
}
  \label{tab:model5}
  \addtolength{\tabcolsep}{6.8pt}
  \begin{tabular}{ccccc} 
    \hline \hline
    $\rho_B^{} / \rho_0^{}$ & $M_l^{*}$  & $m_K^{*}$ & $f_K^{*}$ & $g_{K q q}^{*}$ \\[0.2em] 
    \hline
    $0.00$  & $0.400$ & $0.495$ & $0.091$ & $4.570$ \\ 
    $0.25$ & $0.370$ & $0.465$ & $0.091$ & $4.536$ \\
    $0.50$ & $0.339$ & $0.437$ & $0.090$ & $4.495$ \\
    $0.75$ & $0.307$ & $0.411$ & $0.089$ & $4.455$ \\
    $1.00$ & $0.270$ & $0.386$ & $0.088$ & $4.408$ \\
    $1.25$ & $0.207$ & $0.359$ & $0.084$ & $4.332$
    \\ \hline \hline
  \end{tabular}
\end{table}

\begin{table}[htb]
  \caption{Same as Table~\ref{tab:model5}, but in the QMC-based 
approach ($m_l = 400$ MeV).}
  \label{tab:model2}
  \addtolength{\tabcolsep}{4.8pt}
  \begin{tabular}{ccccc} 
    \hline \hline
    $\rho_B^{} / \rho_0^{}$ & $m_l^*$ &  $m_K^{*}$ & $f_K^{*}$ & $g_{K q q}^{*}$ 
\\[0.2em] 
    \hline
    $0.00$  & $0.400$ & $0.495$ & $0.091$ &  $4.570$ \\ 
    $0.25$ & $0.357$ & $0.452$ & $0.091$ &  $4.520$ \\
    $0.50$ & $0.316$ & $0.418$ & $0.090$ &  $4.467$ \\
    $0.75$ & $0.278$ & $0.390$ & $0.088$ &  $4.419$ \\
    $1.00$ & $0.243$ & $0.371$ & $0.087$ &  $4.376$ \\
    $1.25$ & $0.212$ & $0.360$ & $0.085$ &  $4.339$
    \\ \hline \hline
  \end{tabular}
\end{table}

\begin{figure}[t]
  \centering\includegraphics[width=\columnwidth]{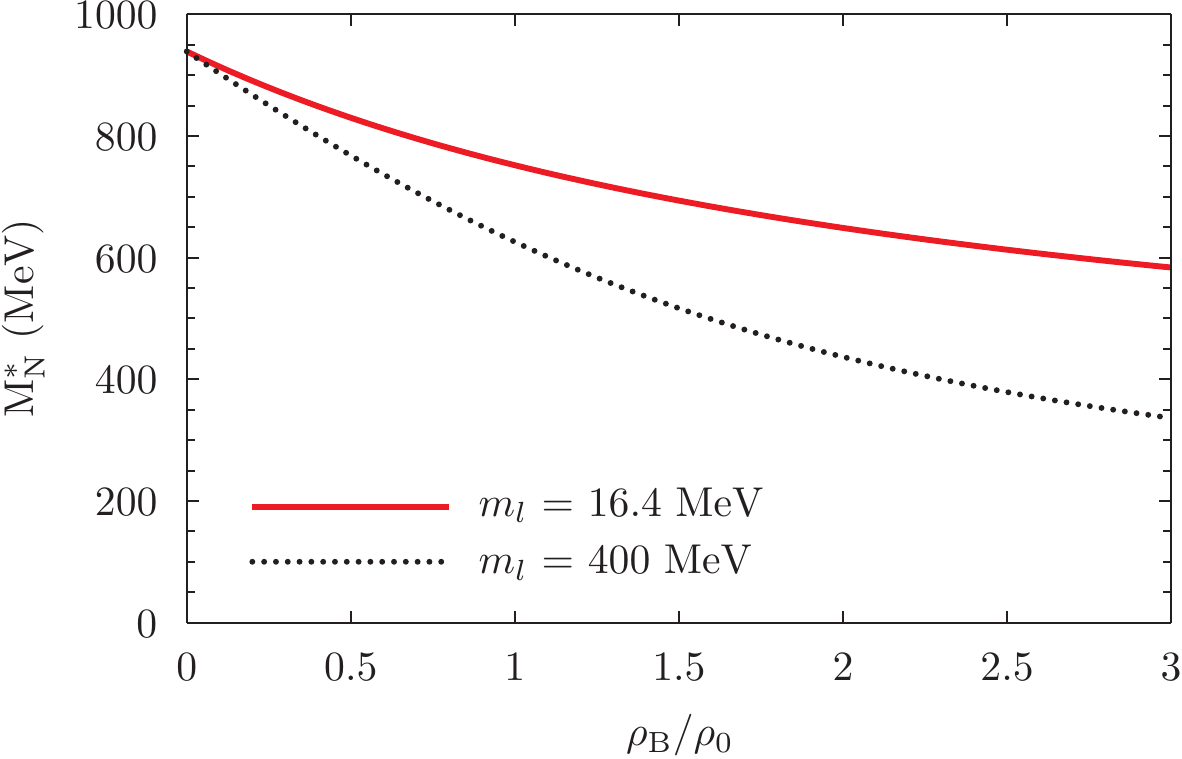}
  \caption{\label{fig2} 
    Effective nucleon mass $M^{*}_N$ for symmetric nuclear matter calculated in the 
    QMC model, corresponding to the NJL-based approach (solid line), and for the QMC-based 
    approach (dotted line).}
\end{figure}

\begin{figure*}[t]
  \centering\includegraphics[width=\textwidth]{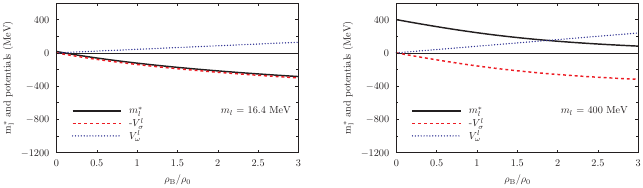}
  \caption{\label{fig3} 
    Effective light-quark mass ($m_l^*$) and quark potentials ($-V^l_{\sigma}$ 
    and $V^l_{\omega}$) used as inputs in the NJL-based approach  
    with $m^{}_l =$ 16.4 MeV (left panel), and those     
    in the QMC-based approach with $m_l^{} =$ 400 MeV (right panel), respectively.
    These are calculated in the QMC model.}
\end{figure*}

Illustrated in Fig.~\ref{fig2} are the effective nucleon mass $M_N^{*}$ 
calculated in the QMC model, corresponding to the two approaches, the NJL-based and the QMC-based. 
The corresponding effective quark mass $m_l^{*}$, scalar $-V^l_\sigma$, 
and vector $V^l_\omega$ 
potentials are shown in Fig.~\ref{fig3}.

\section{$K^+$ Electromagnetic Form Factor in Medium} \label{mediumFormFactors}

In this section, we present the calculation of the $K^+$ EMFF  
in symmetric nuclear matter following Ref.~\cite{HCT16}.
Here, the in-medium light-quark propagators appearing in the Bethe-Salpeter 
equation are modified (see Eq.~(\ref{eq:kaonmed14})).

\begin{figure}[t]
  \centering\includegraphics[width=\columnwidth]{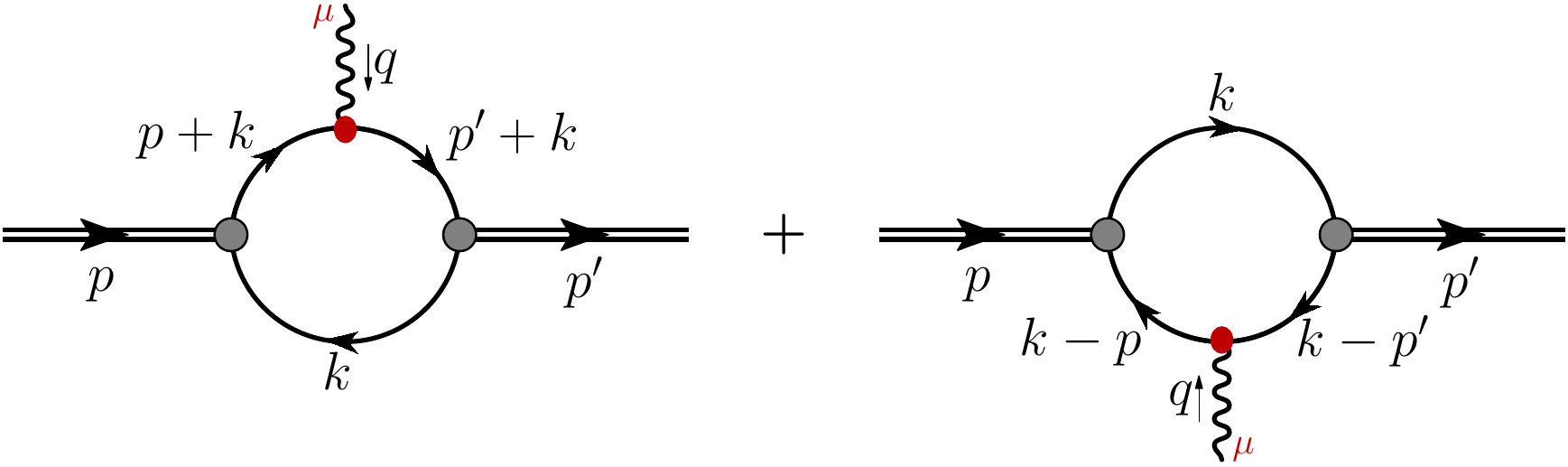}
  \caption{Diagrammatic representation of the electromagnetic interaction for the kaon.}
  \label{fig:emvertex1}
\end{figure}
%

The electromagnetic interaction with the quark is obtained by the minimal substitution:
$i\slashed{\partial} \rightarrow i \slashed{\partial} - \hat{Q}\, A_\mu\,\gamma^\mu$, 
where $A_\mu$ is photon field, 
$e$ is the positron charge and $\hat{Q} = \text{diag}\left[ e_u,\,e_d,\,e_s\right] 
= \frac{e}{2} (\lambda_3 + \frac{1}{\sqrt{3}} \lambda_8)$ is the 
quark charge operator with $e_{u,d,s}$ being the respective quark charges.

The matrix element of the in-medium electromagnetic current for the
$K^+$-meson (will be denoted simply as kaon hereafter, otherwise stated) 
is written as
\begin{align}
  \label{eq:formfactor1}
  J^{\mu} (p'^{*},p^{*}) &= \left(p'^{* \mu} + p^{* \mu} \right) F_K^{*} (Q^2),
\end{align}
where $p^*$ and $p'^{*}$ are the initial and final four-momenta of the kaon, 
respectively, with $q^2 =(p'^{*}-p^{*})^2 \equiv -Q^2$ and $F_K^{*} (Q^2)$ 
denotes the in-medium kaon EMFF.

As the case in vacuum, the in-medium kaon EMFF in the NJL model is calculated 
by the sum of the two Feynman diagrams depicted in Fig.~\ref{fig:emvertex1}, 
which give~\cite{HCT16}
\begin{eqnarray}
  \label{eq:j1}
  \hspace{-4ex}  j^{\mu}_{1,K}\left( p'^{*},p^{*} \right) &=&  i g_{K q q}^{* 2} \int \frac{d^4k}{(2\pi)^4} 
  \nonumber \\ &&\hspace{-15ex} \mbox{} \times
  \mathrm{Tr}\left[ \gamma_5\,\lambda_\alpha^\dagger\,S_l^{*} (p'^{*} 
    + k^*)\,\hat{Q}\,\gamma^\mu\,S_l^{*} (p^{*}+k^*)\,\gamma_5\,\lambda_\alpha\,S_{s}^{*}(k^*)\right], \\
  \label{eq:j2}
  \hspace{-4ex}  j^{\mu}_{2,K}\left( p'^{*}, p^{*} \right) &=&  i g_{K q q}^{* 2} \int \frac{d^4k}{(2\pi)^4}
  \nonumber \\ &&\hspace{-15ex} \mbox{} \times
  \mathrm{Tr}\left[ \gamma_5\,\lambda_\alpha\,S_{s}^{*} (k^*-p^{*})\,\hat{Q}\,
    \gamma^\mu\,S_{s}^{*} (k^*-p'^{*})\,\gamma_5\,\lambda_\alpha^\dagger\,S_l^{*}(k^*)\right],
\end{eqnarray}
where the trace is over the Dirac, color, and flavor indices. 
The index $\alpha$ labels the state and $\lambda_\alpha$ are the corresponding 
flavor matrices. In flavor space the in-medium quark propagator is defined by $S_q^{*} (k^*) 
= \text{diag}[S^{*}_u(k^*),\,S^{*}_d(k^*),\,S^{*}_s(k)=S_s (k)]$ (see Eq.~(\ref{eq:kaonmed14})).

We begin with the relation between the in-medium quark bare EMFFs and kaon EMFF: 
\begin{align}
  \label{eq:bareKplus}
  F_{K^{+}}^{\text{* (bare)}}(Q^2) &= e_u\,f^{* l s}_K(Q^2) - e_s\,f^{* s l}_K(Q^2).
\end{align}
In the expression $f_K^{* ab}(Q^2)$   
the first superscript $a$ indicates the struck quark by the photon, and 
the second superscript $b$ indicates the spectator, and explicit expression is given by 
\begin{eqnarray}
  f^{* ab}_\alpha(Q^2) &=& \frac{3 g_{Kqq}^{* 2}}{4 \pi^2} \int_0^1 dx  \int \frac{d\tau}{\tau}
  \exp \{-\tau [ M_a^{* 2} + x(1-x)\,Q^2 ] \} \nonumber \\ 
  && \hspace{-9ex} + \frac{3 g_{Kqq}^{* 2}}{4 \pi^2}  \int_0^1  dx \int_0^{1-x} dz \int d\tau \nonumber \\ 
  && \hspace{-6ex} \times [(x+z) m_K^{* 2} 
    + (M^{*}_a - M^{*}_b)^2(x+z) + 2\,M^{*}_b ( M^{*}_a - M^{*}_b )] \nonumber \\ 
  && \hspace{-6ex} \times \exp \{-\tau [ (x+z)(x+z-1)\,m_K^{* 2} + (x+z)\,M_a^{* 2} \nonumber \\ 
    && \hspace{15ex} + (1-x-z)\,M_b^{* 2} + x\,z\,Q^2] \}.
\end{eqnarray}
These results are denoted as ``bare'', because the quark-photon vertex 
is elementary, i.e.,  
$\Lambda_{\gamma q}^{\mu\text{(bare)}} = \hat{Q}\,\gamma^\mu$,
with $\hat{Q}$ being the quark-charge operator. 
We note that these expressions satisfy charge conservation exactly.

The quark sector EMFFs in medium for a hadron $\alpha$ can be defined
similarly to those in vacuum~\cite{HCT16}:
\begin{align}
F^{*}_\alpha(Q^2) = e_u\,F_\alpha^{* u} (Q^2) + e_d\,F_\alpha^{* d} (Q^2) 
+ e_s\,F_\alpha^{* s} (Q^2) + \cdots .
\label{eq:quarksector}
\end{align}
The ``bare'' quark form factor in medium $F_\alpha^{* u} (Q^2)$ in
a pseudoscalar meson can easily be related with Eq.~\eqref{eq:bareKplus}.

\begin{figure}[t]
  \centering\includegraphics[width=\columnwidth]{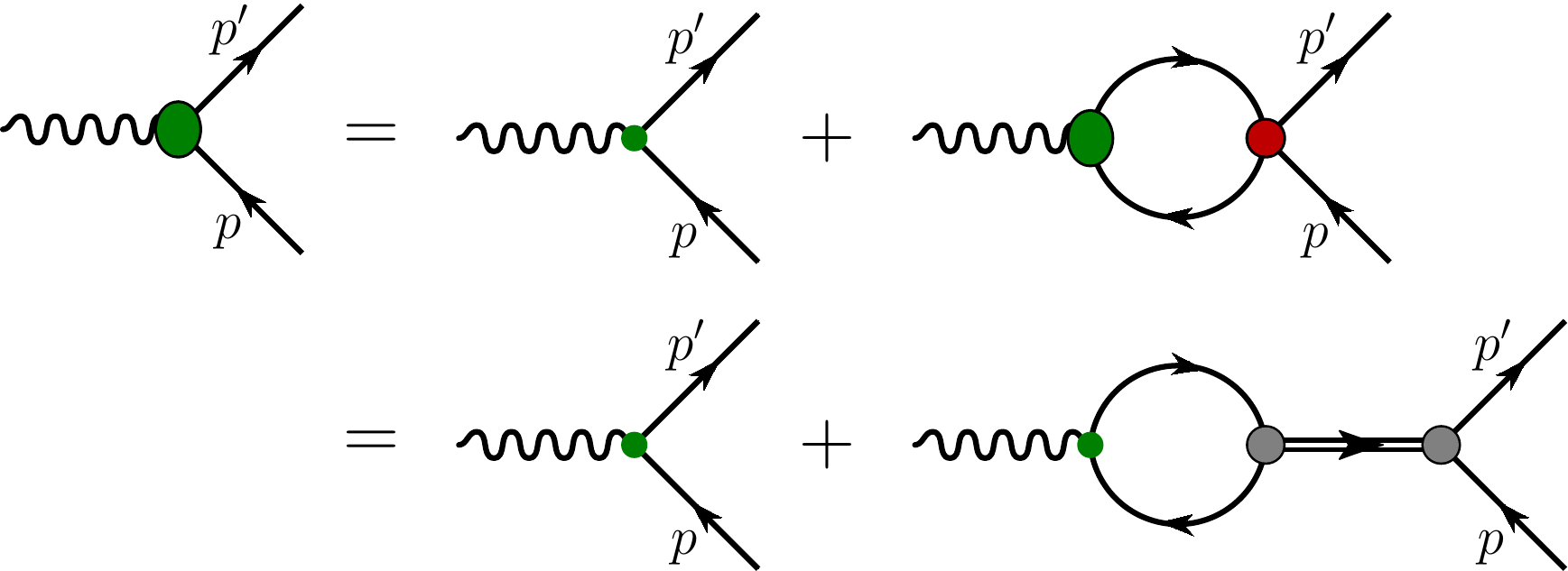}
  \caption{Illustration of the inhomogeneous BSE
    which gives the dressed quark-photon vertex.  
    The large shaded oval represents the solution to the inhomogeneous BSE,
    the small dot is the inhomogeneous driving term ($\hat{Q}\,\gamma^\mu$),
    and the double-dots represents the $q\bar{q}$ interaction kernel~\cite{HCT16}.}
  \label{fig:vectormesons}
\end{figure}

Generally, the quark-photon vertex is not elementary ($\hat{Q}\,\gamma^\mu$) but dressed. 
With the dressing, the quark-photon vertex is given by the inhomogeneous 
Bethe-Salpeter equation as illustrated in Fig.~\ref{fig:vectormesons}.
For denoting the ``dressing'', we replace by $(q,u,d,s) \to (\tilde{Q},U,D,S)$ in the 
rest of this section.
With the NJL-model interaction kernel, the solution to the dressed quark-photon 
vertex for a flavor $\tilde{Q}$ quark has the form: 
\begin{eqnarray}
  \label{eq:quarkvertex}
  \Lambda^{* \mu}_{\gamma\,\tilde{Q}}(p'^{*},p^*) &=& e_q\,\gamma^\mu 
  + \left( \gamma^\mu - \frac{p^{*\mu}\slashed{p}^{*}}{q^{*2}}\right) F^{*}_{\tilde{Q}}(Q^2)
  \nonumber \\ 
  &\to& \gamma^\mu\,F^{*}_{1\tilde{Q}}(Q^2).
\end{eqnarray}
To arrive at the final result, we have used that the $p^{*\mu}\slashed{p}^{*}/p^{*2}$ 
term does not contribute to the electromagnetic current by the current conservation.
Note that this form clearly satisfies the Ward-Takahashi identity 
$q_\mu\,\Lambda^{* \mu}_{\gamma\,\tilde{Q}}(p'^{*},p^{*}) 
= e_{\tilde{Q}} \left[ S_{\tilde{Q}}^{*-1}(p'^{*}) - S_{\tilde{Q}}^{* -1}(p^{*}) \right]$.

For the dressed $l = u, d$, and $s$ quarks, we find
\begin{align}
  \label{eq:f1U}
  F^{*}_{1\, l=U,D}(Q^2) &= \frac{e_{l}^{}}{1 + 2 G_\rho\,\Pi_{v}^{* l l}(Q^2)}, \\
  \label{eq:f1S}
  F^{*}_{1\, S}(Q^2) &= \frac{e_s^{}}{1 + 2 G_\rho\,\Pi_{v}^{* ss}(Q^2)},
\end{align}
where the explicit form of the in-medium bubble diagram is
\begin{eqnarray}
\Pi^{* qq}_{v} (Q^2) &=& \frac{3Q^2}{\pi^2} \int_0^1 dx \int \frac{d\tau}{\tau} x ( 1-x )
\nonumber \\ && \mbox{} \times 
 \exp \left\{-\tau\left[ M_q^{* 2} + x\left( 1-x\right) Q^2\right] \right\}.
\end{eqnarray}

As in the vacuum case~\cite{HCT16}, we neglect the quark 
flavor mixing for the in-medium dressed quark form factors, 
as well as the in-medium dressed quark masses.%
\footnote{In the limit $Q^2 \to \infty$ these form factors reduce to 
the elementary quark charges as expected from asymptotic freedom in QCD. 
For small $Q^2$ these results are similar to the expectations 
from vector meson dominance, where the $u$ and $d$ quarks are dressed by 
$\rho$ and $\omega$ mesons and the $s$ quark by the $\phi$ meson. 
Note, the denominators in Eqs.~\eqref{eq:f1U} and \eqref{eq:f1S} are 
the same as the pole condition obtained by solving the Bethe-Salpeter 
equation in the $\rho$, $\omega$ or $\phi$ channels.
Therefore, the dressed $u$ and $d$ quark form factors have poles 
at $Q^2 = -m_\rho^2 = -m_\omega^2$, while the dressed $s$ quark 
form factor has a pole at $Q^2 = -m_\phi^2$.}

The final expression for the in-medium $K^+$ EMFF with a dressed quark-photon 
vertex is given by 
\begin{align}
  \label{eq:fullKaon}
  F^{*}_{K^{+}}(Q^2) &= F^{*}_{1U}(Q^2)\,f^{* l s}_K(Q^2) - F^{*}_{1S}(Q^2)\,
  f^{* s l}_K(Q^2),
\end{align}
where the in-medium quark EMFFs are obtained by 
Eqs.~\eqref{eq:quarksector}-\eqref{eq:f1S}.
(See also Eq.~(\ref{eq:bareKplus})).

\section{Numerical Results} \label{results}

\begin{figure*}[t]
  \centering\includegraphics[width=\textwidth]{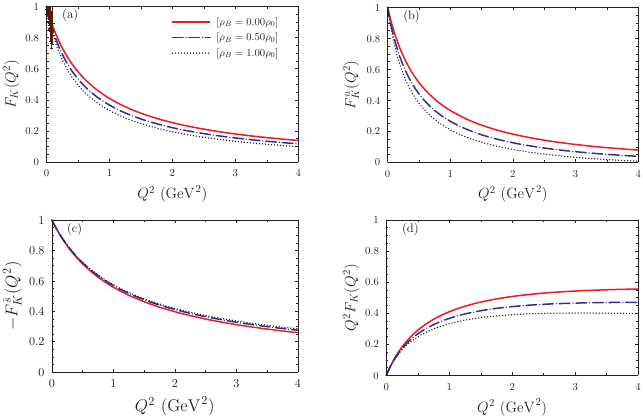}
  \caption{\label{fig7} 
    $K^{+}$ EMFF in symmetric nuclear matter calculated by the QMC-based approach,
    (a) Total kaon EMFF, $F_{K}^{} (Q^2)$ as a function of $Q^2$ for a few densities,
    (b) Up quark EMFF, $F_{K}^{u} (Q^2)$ as a function of $Q^2$ for a few densities,
    (c) Strange quark EMFF, -$F_{K}^{\bar{s}} (Q^2)$ as a function of $Q^2$ 
    for a few densities, and 
    (d) $Q^2 F_K (Q^2)$ as a function of $Q^2$ for a few densities.
    The lines are for $\rho_B^{} / \rho_0^{} =$ 0.00 (solid line), 0.50 (dash-dotted line), and 1.00 (dotted line), respectively.}
\end{figure*}

\begin{figure*}[t]
  \centering\includegraphics[width=\textwidth]{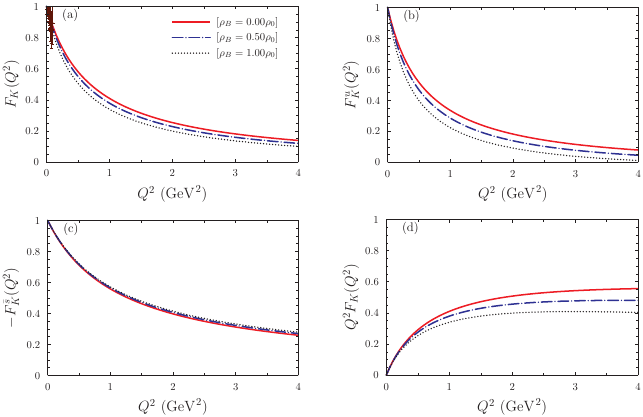}
  \caption{\label{fig8} 
    Same as Fig.~\ref{fig7} but for the NJL-based approach.
    The lines are for $\rho_B^{} / \rho_0^{} =$ 0.00 (dashed line), 0.50 (dash-dotted line), 
    and 1.00 (dotted line), respectively}
\end{figure*}

\begin{figure*}[t]
  \centering\includegraphics[width=\textwidth]{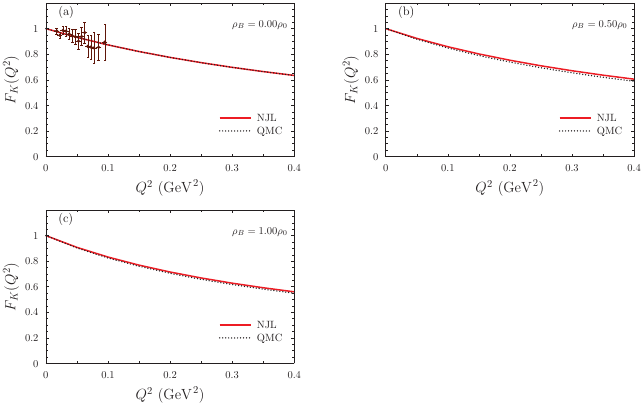}
  \caption{\label{fig9} 
    In-medium $K^+$ EMFF in the QMC-based and the NJL-based approaches,
    compared with the existing data in vacuum ($\rho_B^{} / \rho_0^{} =$ 0.00) at low $Q^2$. 
    Experimental data are in vacuum are from Ref.~\cite{ABBB86b}.}
\end{figure*}

We present our results for the $K^+$-meson EMFF in the two approaches:
the QMC-based and the NJL-based. 
First, we discuss the results obtained by the QMC-based approach.
Those are calculated using the density dependent effective light-quark masses 
determined by the QMC model with the NJL-model fixed  
vacuum values, $m_l =$ 400~$\rm{MeV}$ and $m_s=$ 611~$\rm{MeV}$ (See Table.~\ref{tab:model2}).

The in-medium $K^+$ EMFF along with the $K^+$ 
vacuum form factor and its $u$ and $s$ quark form factors
are shown in Fig.~\ref{fig7}(a)-(d).
Figure~\ref{fig7}(a) shows the $Q^2$ dependence of the in-medium total $K^+$ EMFF 
as well as the in-medium quark EMFFs for a few nuclear densities, 
$\rho_B^{} /\rho_0^{} =$ 0.00 (solid line), 0.50 (dash-dotted line), 
and 1.00 (dotted line), where $\rho_B^{}$ and $\rho_0^{}$ 
are respectively the nuclear density and the normal nuclear 
density, $\rho_0^{} = 0.15$~fm$^{-3}$.

For the vacuum case, we find an excellent agreement with 
the existing data points in Ref.~\cite{ABBB86b} as well as 
the empirical monopole function of $F_K (Q^2) = [1 +Q^2/\Lambda_K^2]^{-1}$~\cite{ABBB86b} 
with $\Lambda_K^2 =$ 0.687 $\textrm{GeV}^2$.
We find the total $K^+$ EMFF, as well as the $u$-quark EMFF in the $K^+$-meson
in medium (shown in Fig.~\ref{fig7}(b)), decrease faster than those corresponding in vacuum 
with $Q^2$ as nuclear density increases.
As in Figs.~\ref{fig7}(b) and~\ref{fig7}(c), we find that the in-medium strange quark EMFF  
begins to dominate the $K^+$ EMFF for $Q^2 \geq 1.6$~GeV$^2$
compared to the $u$-quark EMFF, and
this becomes more pronounced at larger $Q^2$ and higher 
density $\rho_B^{}$. 
The medium effect can clearly be seen, when the $K^+$ EMFF is multiplied by 
the $Q^2$ as shown in Fig.~\ref{fig7}(d).

Next, we discuss the results obtained by the NJL-based approach.
Results for the in-medium $K^+$ EMFF are calculated  
with the NJL model dynamical quark masses in medium, $M_u^{*}$ and $M_s^{*} = M_s$,
using the inputs from the QMC model obtained by $m_l =$ 16.4 MeV 
and $ m_s=$ 356 MeV in vacuum as well as $m^*_l$ in medium.
The kaon EMFF results are shown in Fig.~\ref{fig8}(a)-(d).

Figure~\ref{fig8}(a)-(d) show similar features with those shown 
in Fig.~\ref{fig7}(a)-(d) for both the in-medium $K^+$ EMFF, 
in-medium $u$ and $s$ quark EMFFs in the $K^+$, and $Q^2 F_K(Q^2)$, respectively.
The result of the NJL-based approach shows the similar trend as that of 
the QMC-based approach. This can be understood, since the values of the effective quark masses, 
effective kaon mass, and in-medium kaon-quark coupling constant 
are nearly the same between the two approaches. Based on this fact,
the similar features can also be expected  
for the other mesons by the two approaches 
adopted in this study. For large $Q^2$, the in-medium $K^+$ EMFF for a few nuclear densities 
in both approaches have a good agreement with the QCD result 
in the conformal limit~\cite{LB79} as the vacuum case~\cite{HCT16}. 

In Fig.~\ref{fig9} we show the total $K^+$ EMFF calculated by the 
QMC-based and NJL-based approaches for a few nuclear densities, 
compared with the existing data in vacuum.

\begin{table}[b]
  \caption{$K^+$ charge radius and quark charge radii calculated in the QMC-based approach.
    The results are calculated using the in-medium inputs from the QMC model generated
    for $m_u^{} =$ 400 MeV and $m_s^{}=$ 611 MeV,
    namely the density dependent $m^*_u$ together with $m^*_s = m_s$. 
    All the charge radii are in units of fm.
    The empirical result in vacuum is from Ref.~\cite{PDG18}.}
  \label{tab:model6}
  \addtolength{\tabcolsep}{9.8pt}
  \begin{tabular}{ccccc} 
    \hline \hline
    $\rho_B^{} / \rho_0^{}$ & $r_K^{}$  & $r_u^{}$ & $r_s^{}$ & $r^{\rm expt}$ \\[0.2em] 
    \hline
    $0.00$ & $0.59$ & $0.65$ & $0.44$ & $0.56 \pm 0.03$  \\ 
    $0.25$ & $0.62$ & $0.69$ & $0.44$ & \\
    $0.50$ & $0.65$ & $0.73$ & $0.44$ &  \\
    $0.75$ & $0.68$ & $0.77$ & $0.44$ &  \\
    $1.00$ & $0.71$ & $0.81$ & $0.44$ &  \\
    $1.25$ & $0.74$ & $0.85$ & $0.44$ &
    \\ \hline \hline
  \end{tabular}
\end{table}

\begin{table}[t]
  \caption{
    Same as Table~\ref{tab:model6}, but calculated in the NJL-based approach
    using the in-medium inputs from the QMC model generated for $ m_u = 16.4$~MeV,
    and $m_s^{} = 356$~MeV.
    These $m_u$ and $m_s$ yield the NJL model in-vacuum dynamical 
    (constituent) quark masses $M_u = 400$ and $M_s = 611$ MeV, respectively. 
    Using the QMC model generated $m^*_u$ from $m_u = 16.4$ MeV, the NJL model 
    calculates the density dependent dynamical quark mass $M^*_u$, 
    and uses this $M^*_u$ together with $M^*_s = M_s$.}
  \label{tab:model7}
  \addtolength{\tabcolsep}{9.8pt}
  \begin{tabular}{ccccc} 
    \hline \hline
    $\rho_B / \rho_0$ & $r_K$  & $r_u$ & $r_s$ & $r^{\rm expt}$ \\[0.2em] 
    \hline
    $0.00$ & $0.59$ & $0.65$ & $0.44$ & $0.56 \pm 0.03$  \\ 
    $0.25$ & $0.61$ & $0.69$ & $0.44$ &  \\
    $0.50$ & $0.62$ & $0.69$ & $0.44$ &  \\
    $0.75$ & $0.66$ & $0.74$ & $0.44$ &  \\
    $1.00$ & $0.69$ & $0.79$ & $0.44$ &  \\
    $1.25$ & $0.74$ & $0.86$ & $0.44$ &
    \\ \hline \hline
  \end{tabular}
\end{table}

Results for the $K^+$ charge radius%
\footnote{The charge radius of the $K^+$ meson is calculated using the relation: 
  \\$\langle r_K \rangle = \sqrt{-6 \frac{\partial F_K (Q^2)}{\partial Q^2}\mid_{Q^2 =0}}$.}
for several nuclear densities calculated in the QMC-based and NJL-based approaches are listed 
in Tables~\ref{tab:model6} and~\ref{tab:model7}, respectively.
We find that the charge radius in vacuum has an excellent agreement with 
the empirical data in vacuum~\cite{PDG18}.
Table~\ref{tab:model6} clearly shows that the values of the $u$-quark charge radius 
as well as the kaon charge radius increase as nuclear density increases. 
At normal nuclear density, the kaon and the $u$-quark charge radii 
increase respectively about 20\% and 25\% relative to those corresponding in vacuum. 
For the $s$-quark, the value of the charge radius in medium is identical 
to that in vacuum, because the s-quark properties 
are not modified in medium in the present approach.
Table~\ref{tab:model7} shows that at normal nuclear density $\rho_0$, the $K^+$ and $u$-quark charge radii increase 
respectively about 17\% and 22\% relative to those corresponding in vacuum.

Overall we find the results calculated by both the QMC-based and NJL-based 
approaches show very similar behavior for the in-medium $K^+$ and quark EMFFs 
up to the nuclear density $\rho_B \leq 1.25 \rho_0$, 
where the in-medium constituent quark masses for the both approaches are almost 
the same in this region as in Fig.~\ref{fig:mq*}. 
Consequently, they give nearly the same results for the in-medium 
$K^+$ EMFF and quark charge radii in this density region,  
as can be seen in Tables.~\ref{tab:model6} and~\ref{tab:model7}.
The difference in the in-medium $K^+$ charge radii in the both approaches  
is about 2\% at normal nuclear density.

\section{Summary and Conclusion} \label{summary}

We have studied the kaon properties   
and the space-like $K^+$-meson electromagnetic  
form factor (EMFF) in symmetric nuclear matter 
in the Nambu-Jona-Lasinio (NJL) model.
We take into account the in-medium effects in the dressed quark-photon 
vertex. 
The in-medium constituent light-quark masses  
are calculated by the two different approaches to use in the NJL model.
The results of the two different approaches 
are compared, and they turned out to give very similar predictions.
Namely, the NJL-based approach that the in-medium dynamical 
light-quark mass $M_l^{*} = M_u^{*}$ calculated by the NJL model
using the in-medium input for the current-quark mass $m_l^*$
(in vacuum $m_l = 16.4$ MeV) generated by the quark-meson coupling model, 
yields nearly the same results as those calculated in the QMC-based approach, 
that uses the in-medium constituent quark mass $m_l^{*}$
(in vacuum $m_l = 400$ MeV) calculated in the QMC model.
In both approaches, the strange quark constituent quark masses 
are kept the same as that in vacuum, $M^*_s = M_s = m_s = m_s^* = 611$ MeV.
The feature of yielding nearly the similar predictions in the two approaches,  
holds up to around the normal nuclear densities.

The difference in the two approaches appears only in the higher density region 
(as shown in Fig.~\ref{fig:mq*}).
At normal nuclear density, the values of the constituent light-quark mass 
$m_l^{*}$ calculated in the QMC-based approach is smaller about 10\% 
than that of calculated in the NJL-based approach.

The NJL model has been used as a complementary model to  
the MIT bag model in the variants of the QMC model in Ref.~\cite{WCT16}.
With similar motivation, we have studied the in-medium kaon properties 
in the NJL model using the in-medium inputs calculated by the QMC model.  
These are either using the in-medium current quark properties for the NJL-based approach, 
or the in-medium constituent quark properties for the QMC-based approach
as mentioned above.

By the NJL-based approach we have calculated the effective kaon mass $m^*_K$,  
kaon decay constant, and kaon-quark coupling constant in symmetric nuclear matter.   
We predict that, at normal nuclear density, the constituent quark mass, effective kaon mass,
kaon-quark coupling constant, and kaon leptonic decay constant decrease respectively 
about 33\%, 22\%, 4\%, and 3\% relative to those corresponding in vacuum.

Alternatively, in the QMC-based approach using the effective 
constituent quark mass of the light quark $m_l^*$ directly calculated in the QMC model,
we have calculated the same quantities using the NJL model 
as those calculated in the NJL-based approach.
In this case we predict, at normal nuclear density, 
the constituent light-quark mass, effective kaon mass, kaon-quark coupling constant, 
and kaon lepton decay constant decrease
respectively about 39\%, 25\%, and 4\%, and 4\% compared to those corresponding in vacuum.

Based on the in-medium kaon properties calculated in the both approaches,  
the QMC-based and the NJL-based, we have studied the $K^+$ EMFF   
in symmetric nuclear matter using the NJL model.
We have found that the in-medium $s$-quark EMFF increases as nuclear 
density increases, which is unexpected.
However, the increasing rate against the increase of nuclear density 
is very small. This leads to nearly unmodified in-medium $s$-quark charge
radius relative to that in vacuum, irrespective of nuclear densities studied.

In contrast, the in-medium $u$-quark EMFF as well as that of $K^+$, 
decrease appreciably as nuclear density increases.
This leads to a larger $u$-quark charge radius as well as that of $K^+$ in symmetric 
nuclear matter. At normal nuclear density, the $K^+$ charge radius $r_K^{}$  
increases about 20\% relative to that in vacuum in the QMC-based approach,  
while it yields about 17 \% increase in the NJL-based approach.
These results indicate that the valence $s$-quark gives a significant 
contribution for the $K^+$ EMFF in vacuum as well as in symmetric nuclear matter, 
for larger $Q^2$ and larger nuclear density $\rho_B$.
Our predictions can be verified by experiments   
such as the Compressed Baryonic Matter (CBM) experiment at GSI~\cite{CB99}.

\begin{acknowledgments}

P.T.P.H. and K.T. thank Yongseok Oh for comments on the manuscript
and useful conversations.
K.T. thanks the Asia Pacific Center for Theoretical Physics (APCTP) 
and Kyungpook National University for warm hospitality and excellent supports 
during his visit. 
P.T.P.H. was supported by the Ministry of Science, ICT and Future Planning, 
Gyeongsangbuk-do, and Pohang City. 
The work of K.T. was supported by the Conselho Nacional de Desenvolvimento 
Cient\'{i}fico e Tecnol\'{o}gico (CNPq) 
Process, No.~313063/2018-4, and No.~426150/2018-0, 
and Funda{\c c}\~{a}o de Amparo \`{a} Pesquisa do Estado 
de S\~{a}o Paulo (FAPESP) Process, No.~2019/00763-0, 
and was also part of the projects, Instituto Nacional de Ci\^{e}ncia e 
Tecnologia -- Nuclear Physics and Applications (INCT-FNA), Brazil, 
Process. No.~464898/2014-5, and FAPESP Tem\'{a}tico, Brazil, Process, 
No.~2017/05660-0.
\end{acknowledgments}

\end{document}